# Numerical Seismic Analysis of Railway Soil-Steel Bridges

**Ehssan Hoomaan[1] and Morteza Esmaeili[2]**

[1]PhD Candidate, Department of Civil Engineering, The City College of New York, New York, NY, E-mail: ehoomaan@ccny.cuny.edu

[2]Professor, School of Railway Engineering, Iran University of Science and Technology, Tehran, Iran, E-mail: m_esmaeili@iust.ac.ir

**ABSTRACT:**
Interaction of corrugated steel plates with surrounding soil medium causes the investigation of seismic behavior of soil- steel structures to be complicated. Reviewing the available literature confirms the considerable scarceness in this area. Therefore, this paper studies the seismic behavior of soil-steel railway bridges using FEM dynamic analysis. The finite element two- and three-dimensional models are developed using ABAQUS package by considering four definite spans in time-domain. Many important issues like the effect of 2D and 3D modeling, use of elastic and elastoplastic material models and the thickness of soil cover height were investigated. Another part of the study was devoted to the investigation of the impact of input motion characteristics such as maximum acceleration amplitude and angle earthquake waves approach to the structure on internal forces and consequently the buckling of the structures. Obtained results show that steel arc internal forces in 3D condition are smaller than the 2D condition. Also, elastoplastic behavior reduces internal forces. Increase in input acceleration amplitude and soil cover height yields to increase in internal forces. No regular pattern found for variation of internal forces with respect to the impact angle of input motion.

## INTRODUCTION

Bridges are accounted as key transportation infrastructures which their construction impose considerable costs on road and railway projects. In addition**,** because of their special structural system, they considered to be the most vulnerable element in a roadway or a railway system. One of the suggested structural systems which have achieved more attraction by designers to be used in various practices as culverts, underpasses, overpasses and railway and roadway bridges in recent three decades is soil-steel structures (Esmaeili et al. 2013). One of the important issues related to this kind of structures, is their proper and accurate design against the applied loads and particularly the earthquake effects. Because of embedded nature of these kinds of structures and soil-structure interaction between soil and corrugated steel plates, their seismic analysis and design is known as an important and complicated matter. Reviewing various codes like AASHTO (2017), ASTM (2006), AS/NZS 2041 (2011), BD (2004) and CHBDC (2014) demonstrated some remarkable shortages in definition of seismic loads and their relevant effects on soil steel structures. On the other hand, reviewing the existing researches on the behavior of these structures shows no comprehensive study in this matter. However, many researches can be referred for embedded concrete structures with similar shape. For instance, the Krauthammer and Chen (1986) studied seismic behavior embedded reinforced concrete culverts considering soil-structure interaction using FEM. Wood and Jenkins (2000) modeled a 2D concrete arch using FEM and analyzed it by pseudo-static and spectral methods. They studied effects of surrounding soil stiffness on internal actions of arch. Also, there are some researches which study soil-steel structure dynamic behavior both by using numerical and experimental methods. Esmaeili and Abdulrazagh (2010) studied dynamic behavior of arced railway soil-steel bridges under earthquake loading using FEM by pseudo-static method and claimed that CHBDC equations should be revised for estimating earthquake actions. In experimental area, Ai-lan et al. (2006) by comparison of shaking table test results on a 1/16 scale model of corrugated vinyl chloride resin as corrugated plates surrounded by compacted soil with those obtained by 2D FEM analysis claimed that the maximum soil pressure and internal forces in dynamic and static condition have no significant difference .Lee et al. (2006) compared the obtained seismic analyses results of a corrugated steel tunnel by using CHBDC pseudo-static and dynamic time-history method and shaking table test. Ito et al. (2006) used 2D FEM for simulation of accomplished centrifuge test on a reinforced concreted embedded culvert and consequently proved the adequate capability of numerical method in anticipating dynamic behavior of these kinds of structures.

Review of carried out studies on seismic behavior of soil-steel structures showed their focus on using 2D models for numerical simulation, while the finite width of these structures necessitate their behavior to be investigated by means of three-dimensional models. In the current study firstly a 2D numerical model of an embedded concrete culvert is developed by using ABAQUS software and its results due to seismic excitation are compared with those obtained by Wood and Jenkins (2000). Consequently, based on close agreement between results, validity of the model is proved. Subsequently, by using the validated model and choosing 4 various spans of soil-steel railway bridges, their 2D and 3D numerical models are established. By selecting of Kobe earthquake as input motion firstly a precise comparison are made between results of 2D and 3D numerical models and so the reliability ranges of 2D models is determined. Finally, to demonstrate the model, many sensitivity analyses are accomplished on various parameters like linear and nonlinear behavior of soil, depth of soil cover, amplitude of input earthquake and angle of earthquake waves approach to structure and consequently the internal forces and buckling ratios in structures are derived. The effect of bridge live load and foundation settlement were not studied in this paper (Davis et al. 2018).

# TWO DIMENSIONAL MODELING OF SOIL-STEEL BRIDGES

## Geometry of Model

Using soil-steel bridges as grade separation of roadways and railways becomes more common; consequently, geometry of the bridges is adjusted according to this demand. On the other side, corrugated steel plates are produced in specific spans and heights. Hence, for covering prevailing spans, 4 geometry types are used for numerical modeling. All four arches are semi-circular. Figure 1 shows the general geometry of the model in 2D and 3D states and Figure 2 shows detailed geometry of abutment.

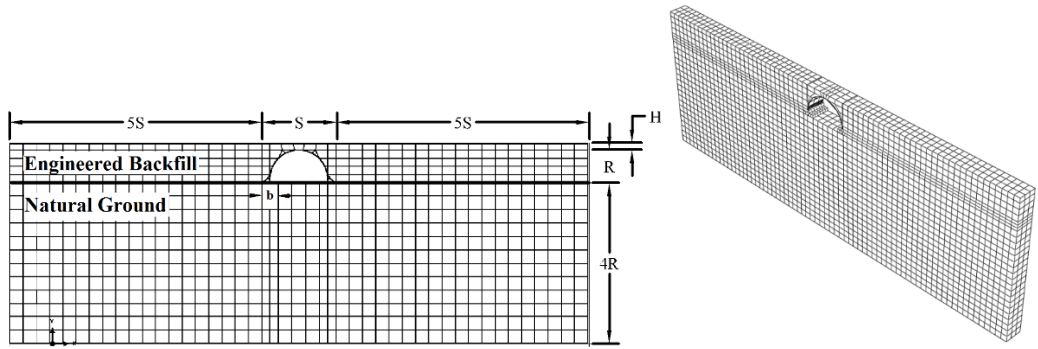

Figure 1. Schematic dimensions of 2D and 3D models and sample of discretization

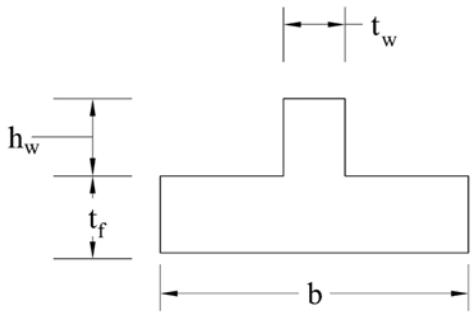
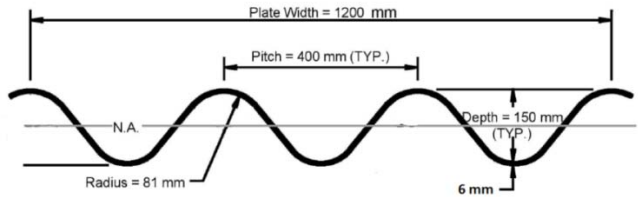

Figure 2. Geometry of abutment

Figure 3. Standard Section of corrugated plate 400 mm × 150 mm with 6 mm thickness (CHBDC 2006)

Figure 3 shows standard corrugated steel plate 400mm × 150 mm × 6 mm which is modeled as a rectangular equivalent section in the software. Geometrical parameters of 4 types soil steel railway bridges are listed in Table 1. The S and R parameter are representing span and rise, respectively. The width and thickness of foundation for first two spans are assumed 3m and 1m, respectively and for second two spans are taken as 4 m and 1.5 m.

Table 1. Geometrical parameter of 4 types of soil-steel railway bridges

| S×R (m×m) | b(m) | $t_f$ (m) | $t_w$ (m) | $h_w$ (m) | H (m) |
|---|---|---|---|---|---|
| 14.13×7.6 | 3 | 1 | 0.8 | 1 | 3.5, 2.5, 1.5 |
| 16.6×8 | 3 | 1 | 0.8 | 1 | 1.5 |
| 20.95×10.47 | 4 | 1.5 | 0.8 | 1 | 1.5 |
| 23.40×11.70 | 4 | 1.5 | 0.8 | 1 | 3.5, 2.5, 1.5 |

**Mechanical Properties of Materials**

In construction of soil-steel bridges, achieving the desire degree of compaction of surrounding soil is the most important part. In this order, proper soil for backfill must consist of course and fine aggregate. Therefore, SC or GC soil type is used for this layer. So, in the present study the properties of the backfill soil are defined based on SC soil. In this matter, Drucker-Prager constitution law is applied for soil behavior. Moduli of elasticity values are taken from CHBDC (2006) according to compaction ratio 90%. In order to consider Young's Modulus variation in depth, properties of soil layer are assumed linear variable with respect to depth. Shear strength parameters are assumed from soil mechanic references and according to used soil type. Whereof Modulus of elasticity is assumed variable with respect to depth, shear strength parameters are also variable in depth. Steel plate behavior is assumed perfect elastoplastic and concrete abutments are modeled as elastic material. The mechanical properties of used materials in 4 selected bridges are listed in Table 2.

Table 2. Mechanical Properties of Material (Geotechnical Engineering Handbook 2002 and CHBDC 2006)

| Mechanical Properties | Unit | Soil Material | | Steel Plate | Concrete Foundation |
|---|---|---|---|---|---|
| | | Soil Cover | Natural Soil | | |
| Mass Density | kg/m³ | 2241.61 | 2241.61 | 7850 | 2500 |
| Module of Elasticity | MPa | 12-24 | 48 | 2×10⁵ | 2×10⁴ |
| Poisson's Ratio | - | 0.3 | 0.3 | 0.3 | 0.3 |
| Friction Angle | Deg | 34-38 | 30 | - | - |
| Cohesion | kPa | 0 | 7 | - | - |
| Yield Stress | MPa | - | - | 300 | - |

Cohesion $d$ and friction angle $\beta$ of Drucker-Prager model are calculated by equation (1) and (2) using friction angle $\phi'$ and Coulomb cohesion $c'$ for 2D plane stain condition (ABAQUS 2010)

$$\tan \beta = \frac{3\sqrt{3} \tan \phi'}{\sqrt{9 + 12 \tan^2 \phi'}} \tag{1}$$

$$d = \frac{3\sqrt{3} c'}{\sqrt{9 + 12 \tan^2 \phi'}} \tag{2}$$

In order to achieve Rayleigh damping coefficients, firstly, for all models, modal analysis are performed and natural frequency of each model is obtained, and then by using equation (3) and according to Table 3, Rayleigh damping coefficients are calculated:

$$\xi_i = \alpha_R / 2\omega_i + \beta_R \omega_i / 2 \tag{3}$$

where, $a_R$ and $\beta_R$ are Rayleigh damping coefficients and $\omega_i$ is natural frequency of $i^{th}$ vibration mode. For example, Table 3 shows these coefficients for arc 14.13m×7.6m with a 3-m soil cover and natural frequency 0.36 rad/s.

Table 3. Damping ratio of materials for a sample soil-steel bridge (Day 2002)

| Material | damping ratio ($\xi$) | $a_R$ | $\beta_R$ |
|---|---|---|---|
| Granular Soil | 0.02 | 0.05 | 0.007 |
| Steel | 0.005 | 0.14 | 0.02 |
| Concrete | 0.05 | 0.01 | 0.001 |

**Model Elements and Boundary Condition**

In 2D model two types of element are used for meshing. Steel plates and abutments are discretized by using a linear 2-nod beam element and for discretizing natural ground and soil cover, a bilinear 4-nod solid element are used. A sample of 2D meshed model is shown in Figure 1.

The model boundary condition is set as follows; the bottom of the model is constrained in vertical direction while in lateral boundaries the spring-dashpots are installed. Springs' stiffness and dashpots' coefficient are calculated by the proposed equations of Makris and Gazetas (1992):

$$c_i = \rho_i A_i V_i \tag{4}$$

$$k_i = K_i A_i \tag{5}$$

$$K_i = 1.2 E_i \tag{6}$$

In above equations $\rho_i$ is mass density of each layer, $A_i$ is mesh area at boundary, $V_i$ is shear wave velocity of each soil layer, $E_i$ is modulus of elasticity of each soil layer and $K_i$ is subgrade modulus. The shear wave velocity is calculated using equation (7) for each soil layer.

$$V_i = \sqrt{G_i / \rho_i} \tag{7}$$

where, $G_i$ is shear modulus of each layer.

**Contact Surfaces**

Because of existence of various components in these types of structures, the contact surface between two different materials should be defined. In modeling process, between outer surface of steel plate and inner surface of soil cover, a surface-to-surface contact with small-sliding formulation is used. In order to define the contact properties, the friction coefficient 0.32 is assigned in tangential direction based on penalty friction formulation while in normal direction; hard contact with penalty constraint enforcement method is used. A surface-based tie constraint is used to make the translational and rotational motion as well as all other active degrees of freedom equal for steel plate and abutment surfaces. Also, these surface-based constraints are used for underneath surface of foundation and natural ground.

**Verification of 2D Numerical Model**

In this section, control of accuracy of 2D modeling procedure is performed by comparison of the results with those reported by Wood and Jenkins (2000). In the mentioned research, an embedded concrete arch was analyzed by FEM in pseudo-static and spectral approaches. The pseudo-static analysis was performed with horizontal acceleration 0.2g and spectral analysis with the Standards New Zealand Loading Code (NZS 1170.5: 2004)elastic spectrum scaled for 20% damping.

Geometry of arch is ellipse with transverse diameter 5.5m and conjugate diameter 5.4m while the concrete arch thickness is 30 cm. Model width and depth nature soil under foundation for soil cover 3m is 40m and 8m, respectively. In the mentioned research, modulus of elasticity, Poisson's ratio and mass density were selected as 20 GPa, 0.2 and 2400 kg/m³, respectively. Comparison of bending moment and shear force in length of concrete arch are shown in Figure 4. As it appears, there is a good agreement between two sets of results.

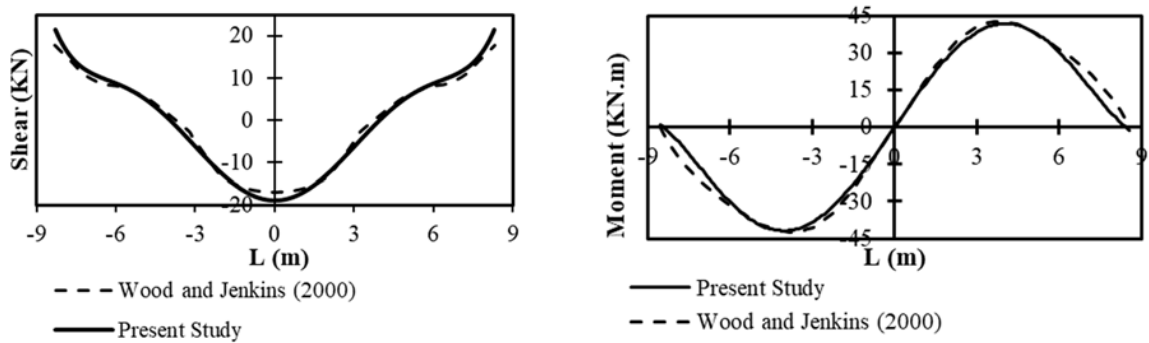

(a) Bending moment　　　　　　　　　　　　(b) Shear force

Figure 4. Diagram of Internal force versus arch length in pseudo-static analysis for (soil cover is 3m)

In the other part of model verification, the results of spectral analysis are compared. To perform spectral analysis, the Standards NZS 4203 elastic spectrum was used, and geometry, boundary conditions, contact surfaces and meshing are the same as pseudo-static analysis. Figure 5 shows the comparison of bending moment for 3m soil cover. Generally, regarding good accordance of obtained results of FEM analysis by using ABAQUS software with those achieved by Wood and Jenkins (2006), the developed model has enough accuracy for investigation of seismic behavior of buried soil-steel structures. In the next section, the process of developing of three dimensional numerical models is explained.

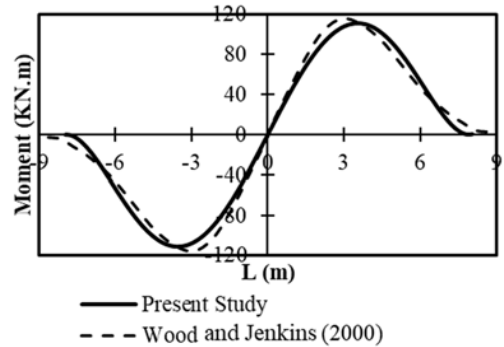

Figure 5. Bending moment diagram versus arch length for in spectral analysis (soil cover is 3m)

## THREE DIMENSIONAL MODELING

### Geometry of Model and Material Properties

In 3D state, the perpendicular dimension to the plane, based on Iran Railway geometric design code (2004), is taken equal to 6 m which is according to one way railway track specifications. All material constitution law and their parameters for steel and concrete are the same as 2D condition. Drucker-Prager parameters are calculated using 3D equations.

### Model Elements and Boundary Conditions

In developed three-dimensional model, bottom of model is constrained in vertical ($y$) and transverse ($z$) directions. In the cases which transverse earthquake component has applied, longitudinal constrained is also considered. Lateral surfaces of model are constrained by spring-dashpot. Spring stiffness and dashpot damping coefficient are considered as 2D model. In 3D state, two kinds of elements are used. For steel plate meshing, S4R element which is a 4-node, quadrilateral, stress/displacement shell element is used. In addition, soil and abutments are discretized using C3D8R which is a solid 8-node linear brick element.

### Loading and Constrains

The loading in seismic analyses of structures in time domain is usually defined by imposing the earthquake acceleration time history on the model base. In this order, longitudinal component of NIS station accelerogram of Kobe (1995) earthquake (Figure 6.a) with PGA 0.51g is selected and applied on the model base in $x$ direction of structure while its transverse component (Figure 6.b) with PGA 0.48g is applied in $z$ direction. In 2D model only the longitudinal component of accelerogram is implemented. In this study, the vertical component of earthquake is neglected. In part of study, to survey the effect of acceleration amplitude, accelerogram is scaled for three acceleration 0.25g, 0.35g and 0.45g.

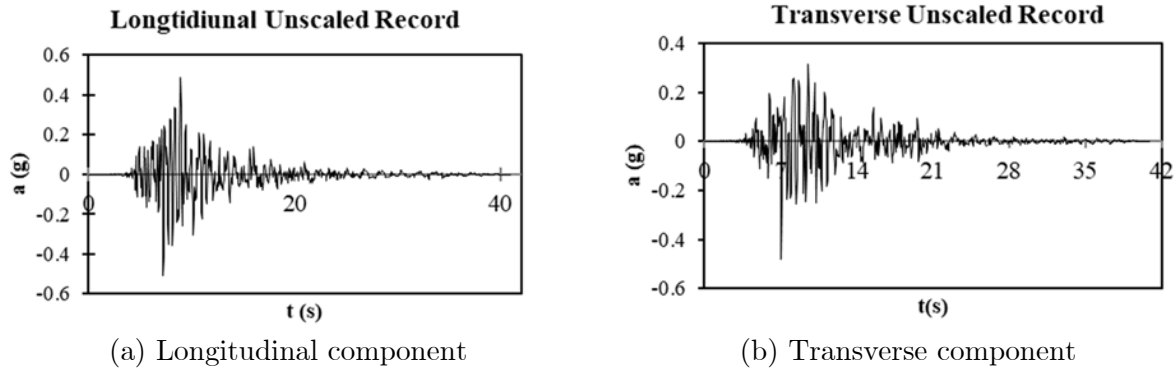

(a) Longitudinal component        (b) Transverse component

Figure 6. NIS station accelerogram of Kobe (1995) earthquake

**Sensitivity Analysis Results**

With regard to mentioned theme in literature review, it was clarified that in seismic behavior of soil-steel bridges, various subjects as predomination of 2D or 3D behavior, shape and geometry of bridge (i.e. arch or box), linear and nonlinear behavior of material and effects of earthquake excitation have not perused. In order to perform a rather comprehensive study through the lens of this viewpoint, following parameters are used for performing sensitivity analysis:

1. Geometry of soil steel bridge structure (4 types of arch structure)
2. Elastic and elasto-plastic behavior of soil
3. Soil cover height
4. Effect of simultaneity of longitudinal and transverse component of earthquake waves
5. Effect of earthquake wave approach to structure
6. The difference between results of 2D and 3D numerical models

**Effect of Material Behavior on Results of 2D and 3D Models**

In order to scrutinize the effects of sensitivity analysis on internal forces, variation diagram of these forces are plotted for 2D and 3D models in linear and non-linear conditions. In 3D models, internal forces values are extracted in three sections of $z = 0$ m, $z = 3$ m and $z = 6$ m. Comparing values in these three sections shows that the first and the third sections' values are the same and greater than the second section. However, axial force variation procedure in medial section is different from two other sections, but its maximum values are smaller. Therefore, in 3D models, internal forces are extracted in the third section $z = 6$ m. All values are recorded in time corresponding to maximum acceleration of input accelerogram.

Figure 7 shows variation of internal forces with respect to span length for 0.25g PGA and for two spans 14.13 m×7.6 m and 23.40 m×11.70 m.

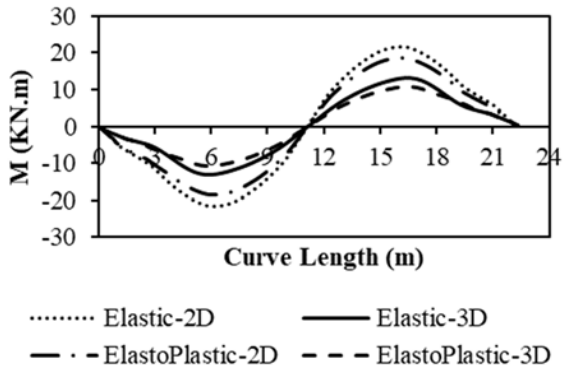
(a) Bending moment for structure type 14.13

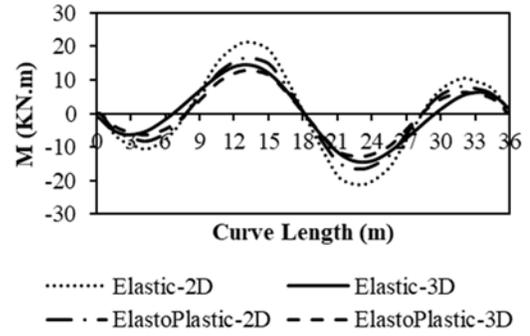
(b) Bending moment for structure type 23.40

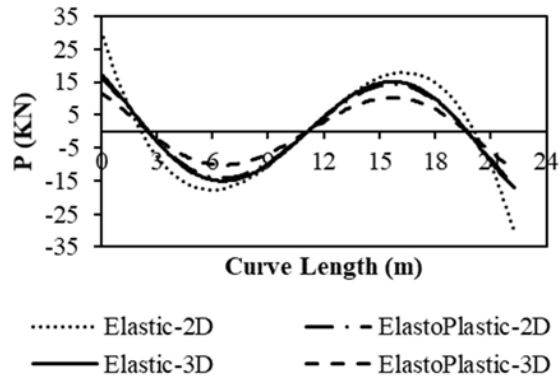
(c) Axial force for structure type 14.13 m×7.6 m

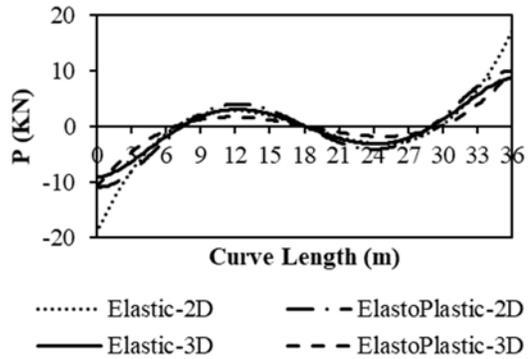
(d) Axial force for structure type 23.40

Figure 7. Internal forces diagrams for PGA 0.25g for structure types 14.13 m×7.6 m and 23.40 m×11.70 m

According to Figure 7 and other diagrams for all structures' dimensions and accelerations, values of internal forces in 3D state with elasto-plastic behavior are smaller than other states. For instances, maximum moments corresponding to PGA 0.25g and for structure type14.13 m×7.6 m in 3D elasto-plastic state is 19% smaller than moment value in 3D elastic model. The relative difference of shear and axial forces are 16% and 43%, respectively. Also, relative difference between elasto-plastic 2D and 3D corresponding to PGA 0.25g for bending moment, shear and axial force are 36%, 23% and 21%, respectively. The mentioned relative differences corresponding to PGA 0.45g for bending moment, shear and axial force are 37%, 22% and 28%, respectively. Hence, as it was observed, except for PGA 0.25g, for two other PGA values maximum difference between elasto-plastic 2D and 3D belongs to maximum bending moment.

This difference can be explained in such way that in 2D state, force distribution in direction of perpendicular to the bridge plane is not possible; as a consequence, one section of steel plate must tolerate all forces, while in 3D state load distribution occurs in lateral direction and various sections in width of bridge will participate in load carrying. On the other side, occurring permanent plastic deformation in elasto-plastic behavior absorbs earthquake energy and consequently decreases internal forces in steel arch.

**Effect of PGA Value on Results of 2D and 3D Models**

In order to scrutinize the pattern of maximum moment variation in steel arch for various spans with respect to different PGAs, diagram of bending moment versus PGA is plotted in Figure 8. The pattern of internal forces in all four states of 2D elastic, 3D elastic, 2D elasto-plastic and 3D elasto-plastic for 0.25g, 0.35g and 0.45g PGAs are similar. The maximum bending moment and shear force are occurred in haunch and crown of arches, respectively.

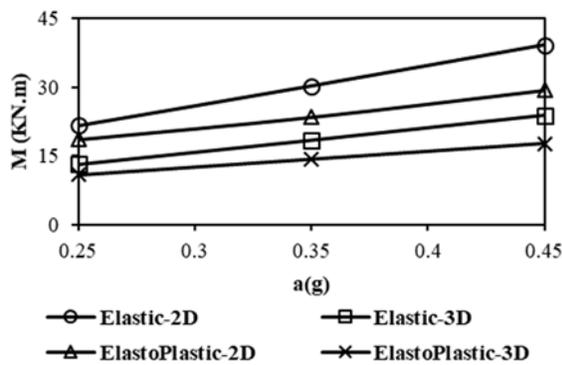
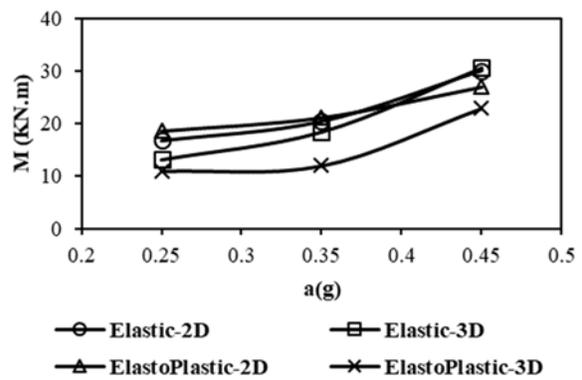

(a) Structure type 14.13 m×7.6 m  
(b) Structure type 16.60 m×8.30 m

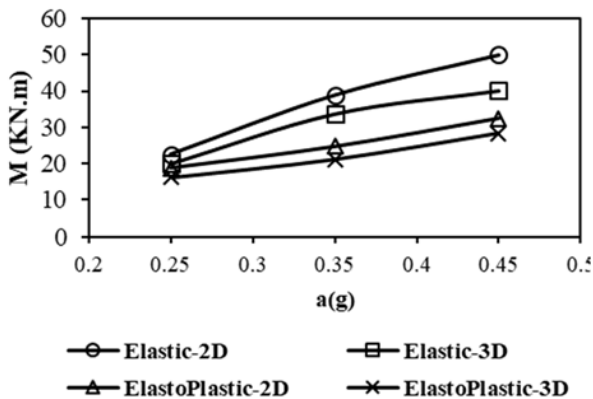
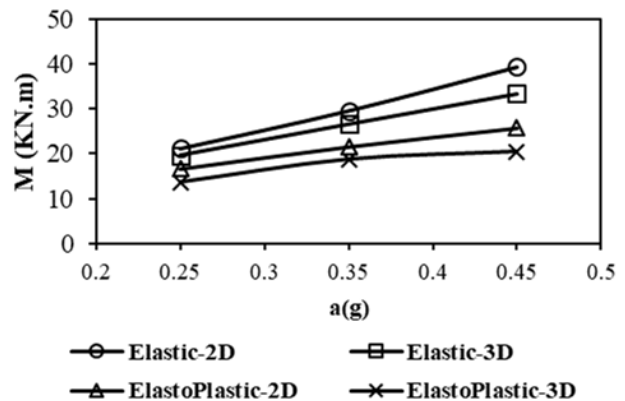

(c) Structure type 20.95 m×10.47 m  
(d) Structure type 23.40 m×11.70 m

Figure 8. Variation of bending moment versus PGA values

As it could be observed in these figures, the pattern of variation for first structure type is linear, but for other types is semi-linear.

# Effect of Earthquake Wave Angle of Approach on Results 3D models

In the real situation, it is possible for a structure to be subjected to earthquake waves in any angle. Whereof in nonlinear behavior of structure and its surrounding soil, the superposition principle is not valid; therefore, decomposition of earthquake force and then applying components in arbitrary direction and finally linear summation of results is not correct approach. Therefore, effect of this phenomenon will be studied in this section. The Kobe earthquake has two longitudinal and transverse accelerogram, in order to change the approaching angle of waves; rotation matrix is used. In order to survey effect of impact angle of earthquake, diagram of internal forces for 3D model and elasto-plastic behavior in 3 sections $z = 0$ m, $z = 3$ m and $z = 6$ m are plotted. Figure 9 shows internal forces with respect to length curve in structure type 14.13 m×7.6m for PGA 0.45g , soil cover 1.5m and for impact angle of 0, 30, 60 and 90 degrees in section $z = 0$ m. Also, Figure 10 shows same parameters in structure type 23.40 m×11.70 m with the same conditions.

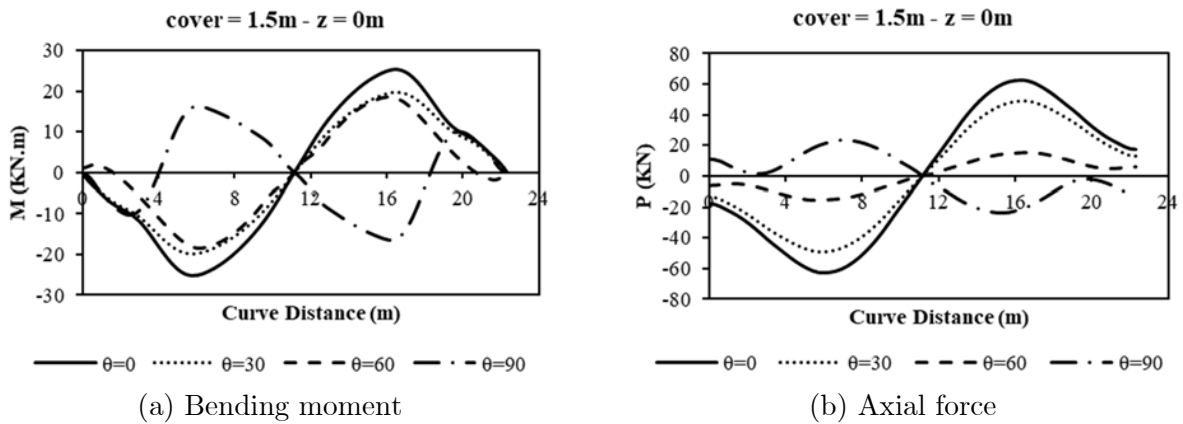

(a) Bending moment        (b) Axial force

Figure 9. Variation of internal forces respect to curve length for 0.45g PGA in structure type 14.13 m ×7.6 m arch

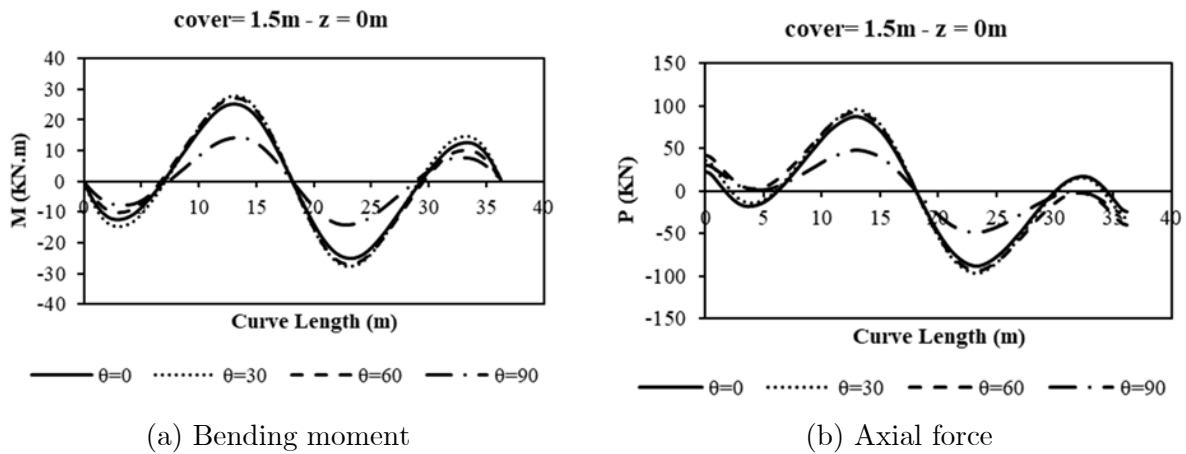

(a) Bending moment        (b) Axial force

Figure 10. Diagram Variation of internal forces with respect to curve length for 0.45g PGA in structure type 23.40 m×11.70 m arch

As it was observed, for soil cover 1.5 m, maximum values of bending moment, shear and axial force are happened in $\vartheta = 0°$, whereas for structure type 23.40 m×11.70 m this angle is 30 degree. Also, it is worth mentioning that in structure type 14.13 m×7.6 m with changing impact angle to 90 degree, pattern of bending moment, shear and axial force changes, since in structure type 23.40 m×11.70 m the values of internal forces are close together. While on the contrary, for soil cover 2.5m and 3.5m these values for different angles change and it is not possible to denote an angle in which internal forces are maximized.

Furthermore, comparison between variation pattern of axial force in different sections in two types of structures shows that for first and last sections ($z = 0$ m and $z = 6$ m) the variation pattern are similar, but, in middle section ($z = 0$ m) this pattern is different, while the variation pattern for bending moment and shear force in all sections are the same In conclusion, from internal force point of view, analysis under impact angle of $\theta = 0°$ yields to critical results.

**The Effect of Soil Cover and Earthquake Approach angle on Results of 3D Models**

In order to compare simultaneous effect of soil cover changes and angle of wave impact on structure on maximum internal forces, results of sensitivity analyses are plotted in Figure 11 and Figure 12. In these figures the variation of internal forces with respect to impact angle for PGA 0.45g and correspond to soil cover 1.5m, 2.5m and 3.5m for two structure type 14.13 m×7.6m and structure type 23.40 m×11.70 m .

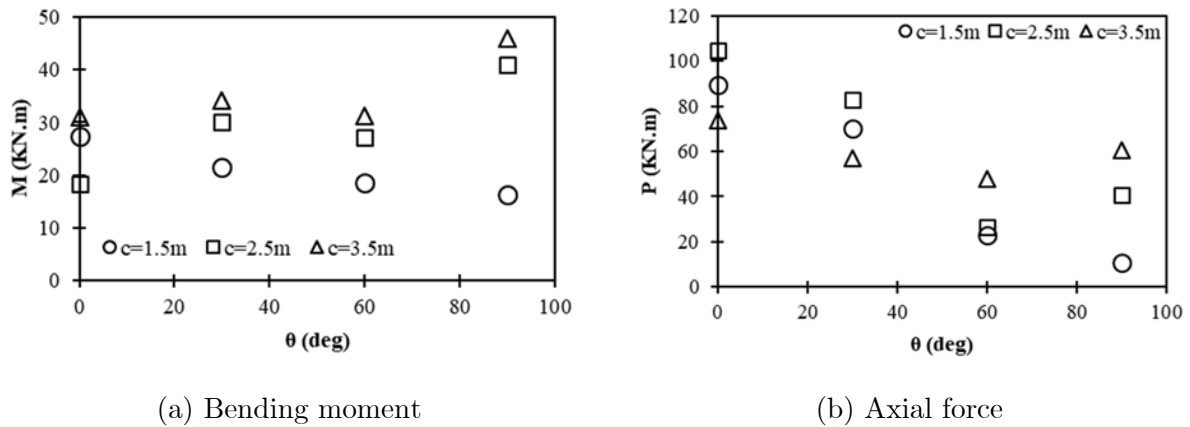

(a) Bending moment                (b) Axial force

Figure 11. Diagram of variation of internal forces variations with respect to impact angle for 0.45g PGA in structure type 14.13 m×7.6 m arch

As it was observed in Figure 11, in structure type 14.13 m×7.6 m arch by increasing earthquake impact angle from 0° to 90° correspond to soil cover 1.5m, maximum bending moment, shear and axial forces decrease, so that moment, shear and axial forces in impact angle of 90° are smaller 40%, 80% and 88% respectively, in comparison with impact angle of zero degree. This pattern for soil cover 2.5m and 3.5m are irregular for all internal forces.

As it is shown in Figure 12 for structure type 23.40 m×11.70 m by increasing in impact angle from 0º to 90º correspond to soil cover 1.5m, maximum bending moment, shear and axial forces at first increase and then decrease so that maximum values of bending moment, shear and axial forces corresponding to $\theta = 90°$ are 43%, 42% and 44%, respectively which are smaller than the case of $\theta = 0°$

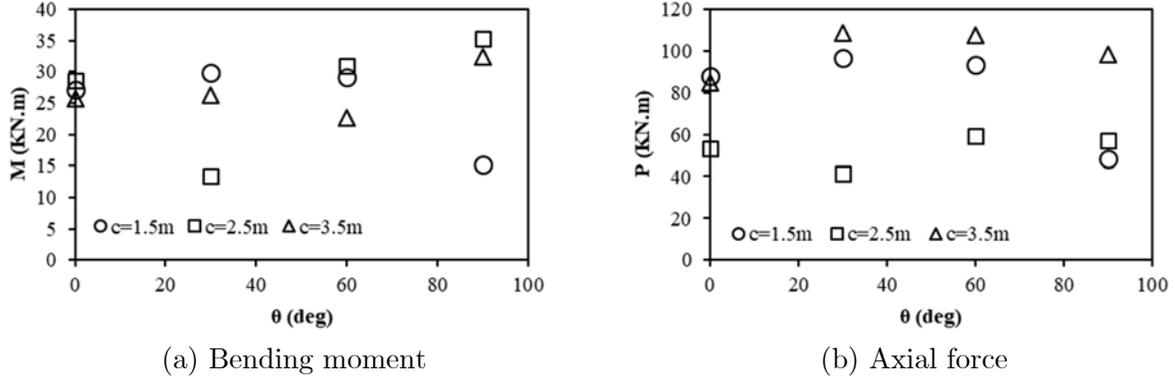

(a) Bending moment  (b) Axial force

Figure 12. Variation of internal forces with respect to impact angle for 0.45g PGA in 23.40 m×11.70 m arch

Also, for soil cover 2.5m this pattern for maximum bending moment is at first decreasing and then increasing, for maximum shear is increasing and for axial force is firstly increasing and then decreasing. For soil cover 3.5m, the pattern of maximum moment and axial force are increasing-decreasing and for maximum shear force is decreasing-increasing.

**Effect of Material Behavior on Buckling Ratio in 2D and 3D Models.**

Required strength for prevention of steel plate buckling in service time is controlled by buckling ratio ($\alpha$). CHBDC (2006) presents following equation for calculating this ratio:

$$\alpha = (P/P_p)^2 + |M/M_p| \leq 1 \tag{10}$$

where in this equation $P$ is axial force, $P_p$ is plastic axial capacity of section, $M$ is bending moment and $M_p$ is bending capacity of section. Table 4 shows plastic properties of section in steel plate.

Table 4. Plastic properties of corrugated steel plate section

| $f_y$ (MPa) | A (mm²/mm) | $P_p$ (kN/m) | Z (mm³/mm) | $M_p$ (kN.m/m) |
|---|---|---|---|---|
| 300 | 8.26 | 2478 | 391.01 | 117.30 |

Figure 13 shows diagram of variation of seismic buckling ratio for quadruplet dimension arch and in 2D and 3D for elastic and elasto-plastic corresponding to PGA 0.45g in curve length. According to Figure 13, buckling ratio in 2D elastic has the maximum, and in 3D elastoplastic has minimum value. In addition, comparison of α values for different PGAs shows that by increasing in PGA, buckling ratio increases. For instance, in structure type 14.13 m×7.6 m for 3D elastoplastic state, α values corresponding to PGA 0.45g are 40% greater than same values corresponding to PGA 0.25g. Also, buckling ratio values in 2D elastoplastic are 37% greater than 3D elastoplastic.

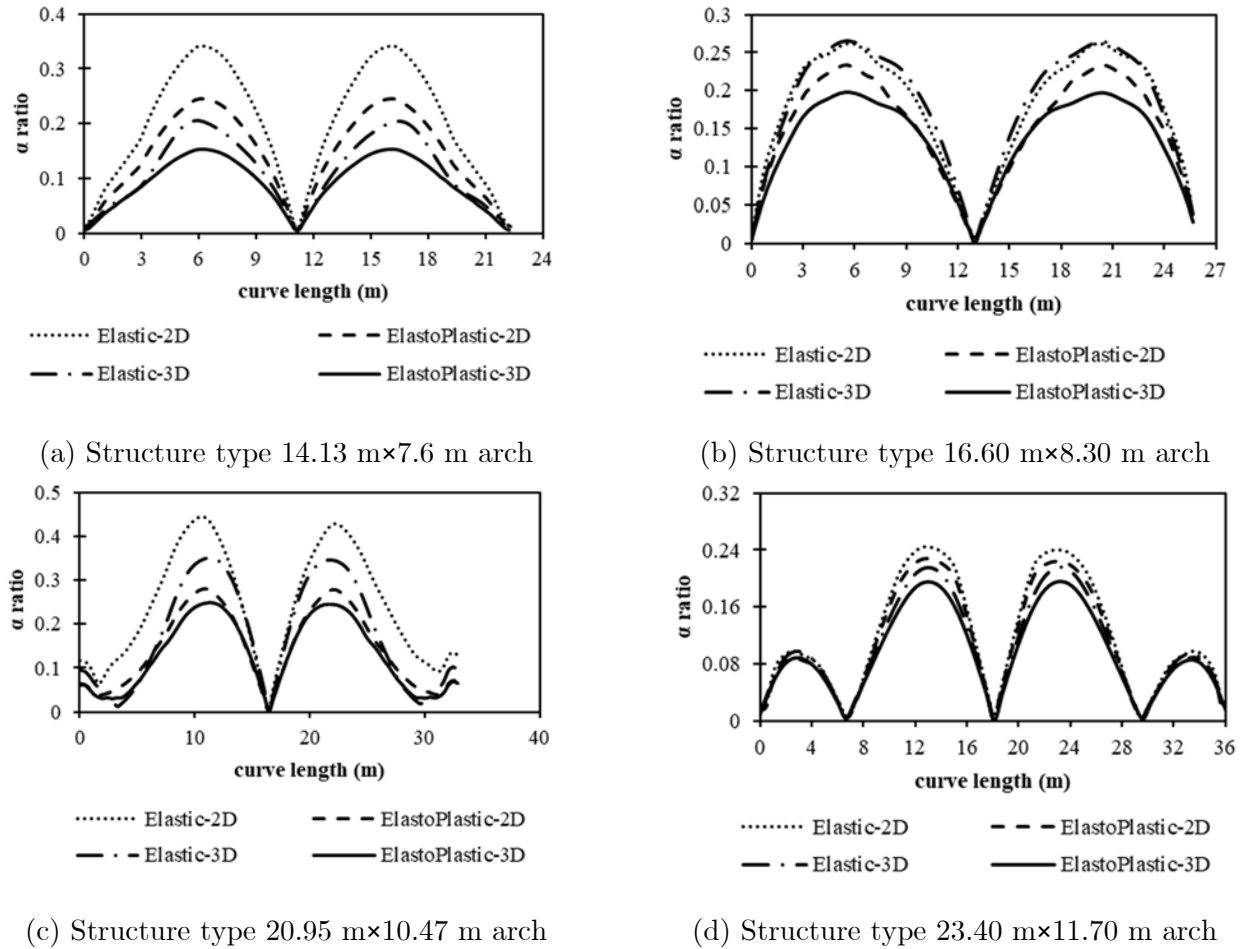

(a) Structure type 14.13 m×7.6 m arch

(b) Structure type 16.60 m×8.30 m arch

(c) Structure type 20.95 m×10.47 m arch

(d) Structure type 23.40 m×11.70 m arch

Figure 13. Variation of buckling ratio along the curve length for PGA 0.45g

In structure type 16.60 m×8.30 m, variation pattern of α coefficient along span length is decreasing-increasing-decreasing and this approach repeats to the end of span. In 3D elasto-plastic state, maximum buckling ratios correspond to PGA 0.45g is 44% smaller than similar values correspond to PGA 0.25g. Maximum α value in 2D elasto-plastic state is 11% greater than 3D elasto-plastic. This process is true for two other spans.

**Effect of Earthquake Impact Angle on Buckling Ratio in 3D Models**

Figure 14 shows variation diagram of buckling ratio of steel arch with respect to earthquake impact angle. As it was observed in Figure 14.a by increasing in impact angle for 14.13m×7.06m arch, buckling ratio for all soil cover decreases. This reduction for soil cover

1.5m is more tangible with respect to other soil covers. By increasing in impact angle from 0º to 90º for soil cover 1.5m, 2.5m and 3.5m, α coefficient decreases 57%, 40% and 39%, respectively. Also, comparison of α values for various soils covers corresponding to an impact angle, shows that soil cover 1.5m and 3.5m have minimum and maximum buckling ratio, respectively.

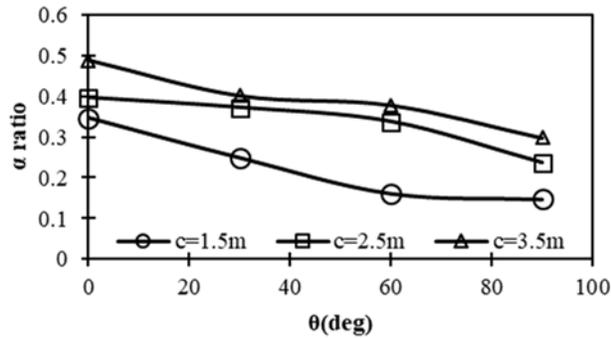
(a) Structure type 14.13 m×7.6 m arch

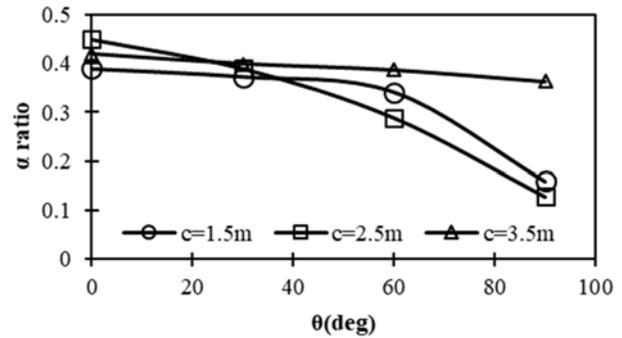
(b) Structure type 23.40 m×11.70 m arch

Figure 14. Variation of buckling ration versus earthquake impact angel

Figure 14.b shows that for 23.40 m×11.70 m arch, increase in impact angle leads to decrease in buckling ratio. Maximum reduction belongs to soil cover 2.5m and equals to 72%. Whereas this reduction for soil cover 1.5m and 3.5m is 59% and 13%, respectively. On the other hand, comparison of α values for various soil covers corresponding to an impact angle shows that unlike previous span, no regular pattern is governing and depending on impact angle the pattern is different. Thereupon, with respect to obtained results, it can be stated that in buckling of steel plate point of view, impact angle of 0º is critical and in comparison to other angles, it gives smaller buckling ratio.

## CONCLUSION

In this paper, the effects of earthquake loading on soil-steel structure were studied. In this order by using finite element method, 2D and 3D model of these kinds of bridges were developed and dynamic analyses were performed under Kobe earthquake input motion. Sensitivity analyses were executed on parameters like soil cover height, structure geometry, PGA value, elastic and elasto-plastic material behavior and 2D and 3D geometry of structures. A summary of most important findings of numerical study can explained as follows:

1. Values of internal forces of 2D models are smaller than corresponding values in 3D models. Maximum values of internal forces for various spans and PGA, elastic and elasto-plastic behavior in 2D condition is 2% to 57% greater than 3D condition.
2. The maximum values of internal forces for various spans and PGAs for both elastic and elasto-plastic behavior, in 2D models are 2% to 57% smaller than 3D models. Also, internal forces values obtained from elasto-plastic analysis are smaller than elastic ones.

3. The maximum values of bending moment for all spans and PGAs, for 2D and 3D models in elastic behavior are 12% to 40% greater than elasto-plastic behavior. These variations for axial and shear forces are 10% to 30% and 12% to 68%, respectively.
4. Increase in soil cover height from 1.5m to 3.5m causes increase in internal forces value of steel plate between 2% to 59%.
5. Generally, by increasing PGA, maximum internal forces increase. This increase is different for various arch dimensions and depends on 2D and 3D modeling and elastic and elasto-plastic behavior. The ranges of these variations are between 30% and 50%.
6. In general, by changing impact angle of earthquake waves from 0º to 90º, no regular pattern found for variation of internal forces.
7. Comparison of buckling ratio variation due to impact angle changes shows that for all spans and soil covers, increase in impact angle yields to decrease in buckling ratio. The range of this reduction is from 13% to 59%.